\documentstyle[prl,aps,epsfig,prb,floats]{revtex}

\begin{document}
\draft 

\title{Renormalization of the Spin-Peierls Transition due to Phonon Dynamics}
\author{S.~Trebst, N.~Elstner and  H.~Monien} 
\address{Physikalisches Institut, Universit\"at Bonn, 
         Nu\ss allee 12, D-53115 Bonn, Germany}
\twocolumn[ \date{\today} \maketitle \widetext

\begin{abstract}
\begin {center}
  \parbox{14cm}{ We report results from a systematic strong-coupling
    expansion of a spin-$\frac{1}{2}$ Heisenberg chain coupled to
    Einstein phonons.  In the non-adiabatic regime ($\hbar \Omega
    \approx J$) this model is used to describe zero temperature
    properties of {CuGeO$_3$}.  The linked cluster expansion allows the
    determination of observables in the thermodynamic limit preserving
    the full lattice dynamics without a truncation of the phononic
    Hilbert space.  In particular, the spin gap and the dispersion of
    the elementary triplet excitation are calculated up to 10$^{\rm
      th}$ order in a dimer expansion.  The magnetic structure factor
    of the ground state is evaluated up to 6$^{\rm th}$ order. We show
    that the spin-phonon coupling leads to a renormalization of the
    elementary triplet dispersion. Surprisingly in the non-adiabatic
    regime a substantial renormalization of the spin gap only sets in
    at much larger couplings than those proposed for {CuGeO$_3$}. The
    ground state magnetic correlations are found to be hardly effected
    by the spin-phonon coupling, but dominated by the frustrating magnetic 
    interaction in the parameter regime relevant for {CuGeO$_3$}.
  }
\end{center}
\end{abstract}
\pacs{\hspace{1.9cm} 
      PACS numbers: 75.10.Jm, 75.40.Mg, 75.50.Ee, 71.38.+i, 63.20.Kr } ]
\narrowtext

A renewed interest for the magnetic properties of one dimensional spin
chains was created by the observation of a spin-Peierls (SP)
transition in the inorganic compound {CuGeO$_3$} \cite{Hase:93}. The SP phase
of {CuGeO$_3$}\ is characterized by an energy gap in the spin triplet
excitation and a dimerization of the lattice along the chain
direction. Experimentally, the nature of the SP transition was
confirmed by inelastic neutron scattering (INS), susceptibility and
X-ray diffraction experiments
\cite{Pouget:94,Harris:94,Nishi:94,Regnault:96,Braden:98,Riera:95,Fabricius:98}.

The SP transition is usually driven by the coupling of spins to the
lattice.  Previous theoretical work treated the spin-phonon coupling
in terms of a {\em static} dimerization \cite{Cross:79}. This approach
does not allow the lattice to adjust to spin fluctuations and can only
be expected to be valid in the adiabatic regime.  In the case of {CuGeO$_3$}\ 
it has been shown that the magnetic energy scales given by the
antiferromagnetic exchange integral $J$ and the phonon frequencies
$\Omega$ are of the same order of magnitude \cite{Braden:98}. Thus, a
realistic model of {CuGeO$_3$}\ has to include dynamical phonons
\cite{Augier:98,Wellein:98,Weisse:99,Bursill:99,Sandvik:99,Kuehne:99}.
For technical reasons only very little is known about the experimental
relevant regime with next-nearest neighbor coupling and intermediate
phonon coupling \cite{Augier:98,Wellein:98,Weisse:99}. 
In this Letter we present numerically reliable
results in the experimentally relevant regime of intermediate phonon
frequencies including a frustrating next-nearest neighbor interaction.

We investigate a spin-$\frac{1}{2}$ Heisenberg chain coupled to
dispersionless Einstein phonons with frequency $\Omega$
\begin{eqnarray}
H = J \sum_i {[(1+g(b_i^{\dagger} + b^{\phantom{\dagger}}_i))
               \vec{S_i}\cdot\vec{S}_{i+1}
              + \alpha\vec{S_i}\cdot\vec{S}_{i+2}]} \nonumber \\
    + \Omega\sum_i b_i^{\dagger}b^{\phantom{\dagger}}_i
\label{H}
\end{eqnarray}
Here, the $b_i^{\dagger}$ and $b_i$ are the local phonon creation and
annihilation operators respectively and $g$ is the spin-phonon
coupling.  The ratio of nearest to a frustrating next-nearest exchange
coupling is given by $\alpha$.

There are two physically independent mechanisms in this model that can
cause a dimerization.  First, without spin-phonon coupling the
frustrating next-nearest neighbor coupling parameterized by $\alpha$
drives a zero temperature quantum phase transition from a spin liquid
to a dimerized phase at a critical value $\alpha_c$. A value of
$\alpha_c = 0.241$ was determined by numerical studies
\cite{Eggert:96}.  In the anti-adiabatic limit the phonons can be
integrated out and the model can be mapped onto a system with a
frustrating next-nearest neighbor interaction \cite{Uhrig:98}. For
$\alpha=0$ a phase transition from a gapless spin fluid to a gapped
dimerized phase occurs at a non-zero value of the spin-phonon
coupling.  These results were confirmed to hold in the non-adiabatic
and adiabatic regimes by numerical studies
\cite{Bursill:99,Weisse:99,Sandvik:99}.  This in particular is a
feature of the dynamical model, since the static model exhibits a
dimerization for all non-vanishing couplings\cite{Cross:79}.

Our numerical approach is based on a linked cluster expansion 
\cite{Gelfand:90,Gelfand:96} which allows to evaluate physical observables
that are additive when the system separates into disconnected parts. It is an
inherent feature of cluster expansions that physical observables are evaluated
in the thermodynamic limit. 
We calculated expansions for the ground state energy, $E_0$, the dispersion of 
the elementary spin triplet excitation $E_{TS}({\bf q})$ including the spin 
gap $\Delta$ and the structure factor $S({\bf q})$. The resulting series were 
analyzed using Pad\'e extrapolation techniques. 

In order to perform a dimer expansion of the Hamiltonian (\ref{H}) we divide
the Hamiltonian as
\[
H = H_0 + H_1
\]
where we shift the phonon operators
\begin{equation}
b_i = \tilde{b}_i - \frac{g}{\Omega}\vec{S_i}\cdot\vec{S}_{i+1}
\label{Trafo}
\end{equation}
which yields an exactly solvable Hamiltonian $H_0$
\begin{eqnarray}
H_0 = J \sum_{i} {[ \vec{S_i}\cdot\vec{S}_{i+1}
                  - \frac{g^2}{\Omega}(\vec{S_i}\cdot\vec{S}_{i+1})^2 ]}
        + \Omega\sum_{i} \tilde{b}_i^{\dagger}\tilde{b}^{\phantom{\dagger}}_i
        \nonumber \\
        + \Omega\sum_{j} b_j^{\dagger}b^{\phantom{\dagger}}_j
\end{eqnarray}
Here the index $i$ runs over the strong dimer bonds while the index $j$ runs 
over the weak bonds between the dimers. 

The eigenstates of the Hamiltonian $H_0$ are described as a product of spin
and phonon states. The spin states are products of local singlet and triplet
dimer states. As a consequence of the transformation (\ref{Trafo}) the phonon
states on the strong dimer bonds are coherent states given by
\begin{equation}
|\tilde{0}_{s/t}\rangle = \exp(-\eta_{s/t}^2/2)\cdot 
                          \exp(\eta_{s/t} b^{\dagger})|0\rangle
\end{equation}
Here $\eta_s = 3g/4\Omega$ corresponds to a spin singlet state and 
$\eta_t = -g/4\Omega$ corresponds to a spin triplet state.
The weak bonds between the dimers are occupied by $n_j$ quantum phonons, 
where $n_j = b_j^{\dagger}b_j^{\phantom{\dagger}}$.

The perturbing Hamiltonian $H_1$ is then given by
\begin{equation}
H_1 = \lambda J \sum_{j} {[ (1+g(b_j^{\dagger} + b^{\phantom{\dagger}}_j))
                            \vec{S_j}\cdot\vec{S}_{j+1}
      + \alpha\vec{S_j}\cdot\vec{S}_{j+2} ]}
\end{equation}
where the index $j$ runs over the weak bonds between the dimers.
The expansion parameter $\lambda$ describes the ratio of the spin-spin coupling
of the weak and strong bonds. The expansion is systematic in $\lambda$ while
the spin-phonon coupling $g$ and the next-nearest neighbor interaction 
$\alpha$ are fixed for each evaluation of a series.

The total Hilbert-space of the Hamiltonian (\ref{H}) is the tensorial product
of the space of the spin configurations and the phononic space. While there is
a finite set of spin configurations for every finite cluster, the Hilbert 
space associated with the phonons is infinite even for a finite cluster.

As a starting point for series expansions we use an initial wavefunction
that describes a dimerized spin phase of spin singlets with a macroscopic 
occupation of phonon states on the dimer bonds and no quantum 
phonons on the bonds between the dimers. Performing a series expansion, only
a {\em limited} number of quantum phonons is created on the weak bonds
between the dimers. The macroscopic occupation numbers of the phonon states
on the strong dimer bonds are tied to the fluctuations of the underlying spin
states. Thus, the overall phononic Hilbert space is finite for a 
series expansion to finite order. In contrast to previous theoretical work
\cite{Augier:98,Wellein:98} we are not forced to truncate the phononic Hilbert 
space.

Series expansions were performed up to 10$^{\rm th}$ order with ten 
contributing clusters. The largest cluster contains eleven dimers or an 
equivalent of 22 spins. The Hilbert space for this cluster has 9,156,836 
contributing states. Calculations were performed on local workstations.

As a closing remark to these technical aspects it is stressed again that our
calculations do not require any finite size scalings.


The antiferromagnetic exchange integral $J$ and the next-nearest neighbor
exchange integral $J'$ are determined from fits to magnetic susceptibility
data. The parameter set $J = 160~K$ and $\alpha = 0.36$ was estimated for {CuGeO$_3$}
\cite{Riera:95,Fabricius:98}. 
The frequencies of the Peierls-active phonon modes were shown to be of the 
order of the magnetic energy scales \cite{Braden:98}, 
$\Omega_1 = 1~J$ and $\Omega_2 = 2~J$. 
Recently, the corresponding coupling constants \cite{Werner:98} 
were proposed to be $g_1 = -0.096$ and $g_2 = 0.16$.
We will investigate our model with these parameters. To make contact to 
previous numerical studies \cite{Wellein:98} a third parameter set with 
$\Omega_3 = 0.3~J$ and $g_3 = 0.11$ is taken into account.

For the spin-$\frac{1}{2}$ Heisenberg chain with static dimerization $\delta$
several  studies of the gap dependency on the static dimerization were 
performed. The corresponding Hamiltonian is given by
\begin{equation}
H' = J_0 \sum_i {[(1+(-1)^i\delta))\vec{S_i}\cdot\vec{S}_{i+1}
              + \alpha_0\vec{S_i}\cdot\vec{S}_{i+2}]}     
\label{H:static}
\end{equation}
The parameters of this model are related to those of Hamiltonian (\ref{H}) by
\begin{eqnarray}
  J &=& J_0(1+\delta)               \nonumber \\
  \lambda &=& (1-\delta)/(1+\delta) \nonumber \\
  \alpha &=& \alpha_0/(1+\delta) \ .
\label{H:parameters} 
\end{eqnarray}
At the critical point the triplet gap $\Delta$ is known to show a 
$\Delta(\delta) \propto \delta^{2/3}$ dependency \cite{Cross:79}.
For supercritical frustration ($\alpha > \alpha_c$) a dependency 
$\Delta(\delta) - \Delta(0) \propto \delta^{2/3}$ was recently proposed by 
Uhrig et. al. \cite{Uhrig:99}.

In order to recover the original Hamiltonian (\ref{H}) the evaluated series 
have to be analyzed at $\lambda = 1$. We have performed two types of Pad\'e 
approximations. First, motivated by the proposed supercritical dependency, we
evaluate biased Pad\'es fixing the exponent to be $2/3$. Second, assuming a
power law dependency without restricting the exponent, we evaluate Dlog Pad\'e 
approximants \cite{Baker:90} and reintegrated the obtained series to evaluate 
the gap. 

For small supercritical values of $\alpha$ and small coupling constants $g$
both Pad\'e approximants show well matching results.
Varying the degree of numerator and denominator polynoms does not 
substantially change the resulting values justifying both approaches.
For values of $\alpha$ and $g$ far away from critical points the biased Pad\'e
approximants become less reliable while the Dlog Pad\'e approximants still
give stable results as can be seen in table \ref{GapTable}.

\begin{figure}[t]
 \begin{center} 
 \epsfig{file=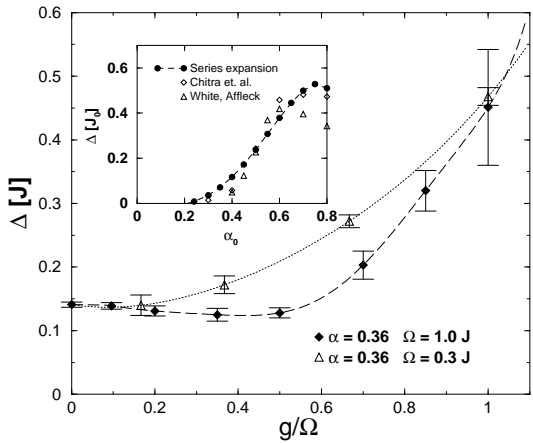, height=6cm}
 \caption[]{Gap versus spin-phonon coupling. The dashed line is a polynomial
            fit to the data.
            The inset shows the gap versus the frustration in the static
            model, filled symbols are obtained by evaluation of the biased 
            $[5,5]$-Pad\'e approximant. The open symbols show DMRG data
            \cite{ChitraWhite:9596}. }
 \label{Gap} \end{center}
\end{figure}

The inset in Fig. \ref{Gap} shows the gap in units of $J_0$ versus the 
frustration $\alpha_0$. The filled symbols are obtained by evaluation of
the biased $[5,5]$-Pad\'e approximant. The open symbols show DMRG data 
\cite{ChitraWhite:9596}.
Our series results are in full agreement with those obtained by the 
perturbative approach using flow equations \cite{Knetter:99}. Good agreement is
also found with previous DMRG results.

Fig. \ref{Gap} shows the gap in units of $J$ for a frustration of 
$\alpha = 0.36$ for various spin-phonon couplings and phonon frequencies. 
The error bars represent the deviation of the accepted Dlog Pad\'e 
approximants. The dashed lines are polynomial fits to the data.
For the parameter set $\alpha = 0.36$, $\Omega_3 = 0.3~J$, and  $g_3 = 0.11$
we find a triplet gap $\Delta = (0.172\pm 0.014)~J$, which is in good agreement
with previous exact diagonalization extrapolations for the $\pi$ phonon mode 
\cite{Wellein:98}.
For the parameters relevant to {CuGeO$_3$}\ our results are below the 
experimentally obtained value $\Delta \approx 0.152~J$ (2.1 meV)
\cite{Harris:94,Nishi:94,Regnault:96}.

In the case of a nearly adiabatic phonon frequency $\Omega = 0.3~J$ the gap
$\Delta$ monotonically increases with higher spin-phonon couplings. For the
non-adiabatic phonon frequency $\Omega = 1~J$ the gap shows a gentle downward
slope. This indicates a suppression of the frustration for small phonon
frequencies due to a stronger renormalization of the nearest neighbor spin
interaction than the next-nearest neighbor spin interaction \cite{Weisse:99}. 
For higher couplings this is compensated by an overcritical coupling.

\begin{figure}[t]
 \begin{center} 
 \epsfig{file=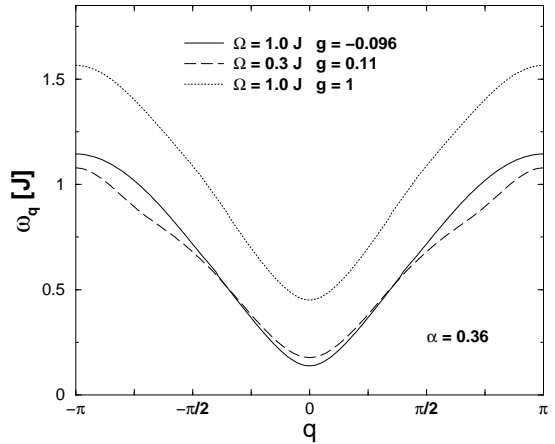, height=6cm}
 \caption[]{Dispersion of the elementary triplet excitation evaluated by a 
            dimer expansion up to 10$^{\rm th}$ order for different phonon 
            frequencies and spin-phonon coupling constants. For each momentum 
            the series were analyzed using Dlog Pad\'e approximants.}
 \label{Dispersion} 
 \end{center}
\end{figure}

The dispersion of the elementary triplet excitation is shown in 
Fig. \ref{Dispersion}. 
The series expansion method is based on Rayleigh-Schr\"odinger perturbation
theory and is therefore supposed to produce well converging results if the 
contributing states are clearly separated by an energy gap. 
In the region $|{\bf q}| > 0.4~\pi$ the series converges very fast. For smaller
momenta Dlog Pad\'e approximants are required to obtain numerically reliable
results.

It turns out that the two parameter sets relevant for {CuGeO$_3$}\ give nearly 
matching dispersion relations. The maxima of the dispersion relation at 
${\bf q}=\pm\pi$ are suppressed in comparison with the known value 
$\omega({\bf q}=\pm\pi) = \pi/2$ for the unperturbed Heisenberg chain which is
mainly due to the next-nearest neighbor magnetic interaction. 


The Peierls ordering structure of the ground state is accompanied by short 
range magnetic order. Thus, we evaluate the following real-space magnetic 
correlation function
\begin{equation}
 S_{j} = \langle \vec{N}_0\cdot\vec{N}_{2j+2} \rangle
\end{equation}
where $\vec{N}_{j}=\vec{S}_{j}-\vec{S}_{j+1}$ is the sublattice 
magnetization 
and $\vec{S}_j$ denotes the spin on site $j$. 
Technically, this is done by formally adding a source term
\begin{equation}
H_2 = h \sum_j \vec{N}_0\cdot\vec{N}_{2j+2}
\end{equation}
to the Hamiltonian (\ref{H}) and expanding the ground state energy to linear 
order in $h$. Differentiating with respect to $h$ will give the strong 
coupling expansion for the correlation function. The determined structure 
factors are well converged. The results presented in the following are 
obtained by direct evaluation of the series without using Pad\'e approximants.

\begin{figure}[t]
 \begin{center} \epsfig{file=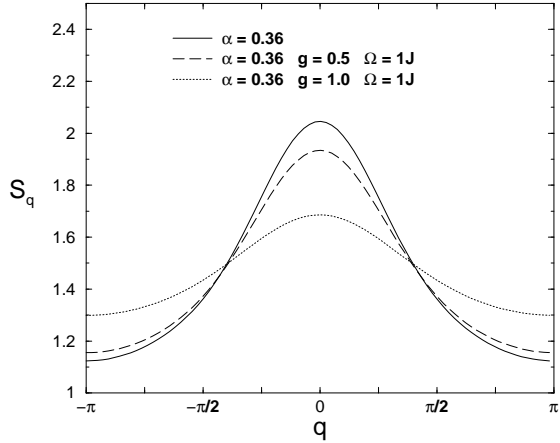, height=6cm}
 \caption[]{Structure factor $S({\bf q})$ versus momentum ${\bf q}$ in a sixth
            order calculation.}
 \label{Sq} \end{center}
\end{figure}

The Fourier transform 
\begin{equation}
S({\bf q}) = \sum_j e^{i {\bf q} \cdot {\bf r}_j} \, S_j
\end{equation}
is shown in Fig. \ref{Sq}.
Parameters are chosen in the gapped phase where the structure factor shows 
a Lorentzian shape which obviously differs from the logarithmic divergence of 
the isotropic Heisenberg model at $S({\bf q}=0)$.
In the case of weak spin-phonon coupling as given in {CuGeO$_3$}\ this is mostly due 
to the next-nearest neighbor interaction. The evaluated series with 
next-nearest neighbor interaction and the series with additional spin-phonon 
coupling cannot be reasonably distinguished.
For stronger couplings Fig. \ref{Sq} shows that the spin-phonon interaction 
leads to a further diminishing of the maximum at ${\bf q} = 0$ and a decrease 
of the spin correlation length.

In conclusion, series expansion techniques were applied to investigate zero
temperature properties of a spin-$\frac{1}{2}$ Heisenberg chain coupled to 
Einstein phonons which in the non-adiabatic limit serves as a model for {CuGeO$_3$}.
We evaluated the dispersion of the elementary triplet excitation and 
demonstrated that the spin-phonon coupling leads to a renormalization of the
dispersion. The dependency of the triplet gap on the spin-phonon coupling shows
a non-monotonous behavior in the non-adiabatic regime. For the parameters 
relevant to {CuGeO$_3$}\ no substantial renormalization of the triplet gap is found.
The magnetic ordering of the ground state was determined by calculation of
the spin structure factor. It was found that the next-nearest neighbor 
interaction dominates the spin correlations in the case of {CuGeO$_3$}.

We acknowledge helpful discussions on this problem with G.~S.~Uhrig, S.~Eggert,
R.~Werner and C.~Gros.



\begin{table}
 \caption{[$n/m$] Dlog Pad\'e approximants to the series for the energy gap 
          $\Delta$ for a frustration $\alpha = 0.36$ and a phonon frequency
          $\Omega = 1~J$. Values are given in units of $J$.}
 \label{GapTable}
 \begin{tabular}{cllll}
     & \multicolumn{1}{c}{[(n-2)/n]} & \multicolumn{1}{c}{[(n-1)/n]}      
     & \multicolumn{1}{c}{[n/n]}     & \multicolumn{1}{c}{[(n+1)/n]}  \\ 
 \hline
 \multicolumn{5}{c}{$g = 0$} \\ 
 n=4 &    -       &  0.136     &  0.145     &  0.099    \\ 
 n=5 &  0.145     &  0.138     &            &           \\ 
 \hline 
 \multicolumn{5}{c}{$g = 0.096$} \\ 
 n=4 &    -       &  0.133     &  0.143     &  0.098    \\ 
 n=5 &  0.143     &  0.135     &            &           \\ 
 \hline 
 \multicolumn{5}{c}{$g = 0.2$} \\ 
 n=4 &    -       &  0.120     &  0.139     &  0.096    \\ 
 n=5 &  0.140     &  0.126     &            &           \\ 
 \hline 
 \multicolumn{5}{c}{$g = 0.5$} \\ 
 n=4 &    -       &    -       &  0.125     &  0.121    \\ 
 n=5 &  0.142     &  0.124     &            &           \\ 
 \hline 
 \multicolumn{5}{c}{$g = 1.0$} \\ 
 n=4 &    -       &  0.277     &  0.461     &  0.503    \\ 
 n=5 &  0.474     &  0.540     &            &           \\
 \end{tabular}
\end{table}

\end{document}